\documentclass{elsart}

\usepackage{natbib}
\usepackage{graphicx}

\usepackage{amssymb}
\usepackage{amsmath}
\usepackage{bm}

\def\url#1{{\ttfamily\def\/{/\discretionary{}{}{}}#1}}
\def\bibcode#1{}

\begin{document}

\begin{frontmatter}
\title{Simulations of Weak Gravitational Lensing}
\author[address1,address2]{Martin White\thanksref{mwemail}},
\author[address2]{Chris Vale\thanksref{cvemail}}
\address[address1]{Department of Astronomy, University of California,
Berkeley, CA, 94720}
\address[address2]{Department of Physics, University of California,
Berkeley, CA, 94720}
\thanks[mwemail]{E-mail: mwhite@astro.berkeley.edu}
\thanks[cvemail]{E-mail: cvale@astro.berkeley.edu}

\begin{abstract}
We describe the simulation data produced by a pilot programme to compute
mock weak gravitational lensing maps for a range of currently popular
cosmological models by ray tracing through high-resolution N-body simulations.
The programme required only a modest investment in computer time to produce
maps accurate to arcminute scales covering hundreds of square degrees of sky
for 4 cosmological models.

\end{abstract}

\begin{keyword}
Cosmology \sep Lensing \sep Large-Scale structures
\PACS 98.65.Dx \sep 98.80.Es \sep 98.70.Vc
\end{keyword}
\end{frontmatter}

\section{Introduction}

Weak gravitational lensing by large-scale structure has become an
indispensable tool for modern cosmology, used already to set constraints
on the mass density ($\Omega_{\rm mat}$, in units of the critical density)
and the fluctuation amplitude $(\sigma_8)$
\citep[see e.g.][for the current status]{WaeMel,HoeYeeGla} 
and touted for its potential to constrain cluster scaling relations
\citep{HutWhi} and dark energy
\citep{BenBer,Hut02,Hu02,Hea03,AbaDod,Ref03,JaiTay,BerJai,TakJai03,TakWhi}.
Like the anisotropies in the cosmic microwave background (CMB), the theory
of weak gravitational lensing is well understood 
\citep{LensReview1,LensReview2}.
Unlike the CMB however the calculation involves modeling the non-linear
evolution of the mass in the universe.
This makes the predictions of weak lensing very rich, but also means that
an accurate treatment requires numerical simulations.

In this paper we give some details of a pilot program designed to provide
weak lensing ray-tracing simulations of a small grid of cosmological models
in the currently favored region of parameter space.
These models can be used by observers wishing to test their algorithms or
fit to their data and by theorists wishing to test or calibrate fast and
flexible, but approximate, methods of calculation.
The initial grid is small, but the process of its creation is almost
entirely automatic allowing it to be expanded simply as the need arises.

This paper focuses on the creation of the model grid.
We describe the choice of cosmological parameters in \S\ref{sec:param},
the N-body simulations in \S\ref{sec:nbody} and the ray tracing and
numerical issues in \S\ref{sec:lensing} and \S\ref{sec:numerics}.
We give some representative results in \S\ref{sec:results} before
concluding in \S\ref{sec:conclusions}.

\section{Cosmological models}

\subsection{Choosing parameters} \label{sec:param}

We choose a small number of cosmological models which provide good fits to
the current CMB and large-scale structure data.  For simplicity all of the
models in the pilot program are variants of the cold dark matter model with
a dark energy component, and their parameters are shown in
Table \ref{tab:model}.
The parameter variations around a base model are designed to keep the CMB
fluctuations almost invariant, in anticipation of increasingly precise
data from WMAP and Planck, and the entire process is automated using Perl
scripts and C code.

We begin with a relatively small number of models to demonstrate the cost
and feasibility of making such grids.  As we gain experience in using these
data products and understand the drivers better we can extend the model
grid and/or increase the fidelity of the simulations.  Are our priorities
to have larger maps?  higher resolution?  more redshift range?  more `sky'?
etc.

\begin{table}
\begin{center}
\begin{tabular}{cccccccc}
Model & $\Omega_{\rm mat}$ & $w$ & $h$ & $n$ & $\sigma_8$ & $\tau$ &
 $\chi^2$ \\  \hline
  1\&2   &  0.296 & -1.0 & 0.70 & 1.00 & 0.93 & 0.15 & 977 \\
  3\&4   &  0.357 & -0.8 & 0.64 & 1.00 & 0.88 & 0.15 & 975 \\
  5\&6   &  0.296 & -1.0 & 0.70 & 0.95 & 0.85 & 0.10 & 979 \\
  7\&8   &  0.357 & -0.8 & 0.64 & 0.95 & 0.81 & 0.10 & 976
\end{tabular}
\end{center}
\caption{Parameters for the models run.  For each cosmological model two
independent sets of initial conditions are generated.  All models are
spatially flat, so $\Omega_{\rm de}=1-\Omega_{\rm mat}$.  For all models
the matter density is $\omega_{\rm mat}=0.145$ and the baryon density
$\omega_b\equiv\Omega_bh^2$ was fixed at $0.023$.  Our results are very
insensitive to this latter choice.
All models have power-law spectra (no running) with index $n$ and the dark
energy has a constant equation of state $w$.  The normalization, $\sigma_8$,
is from a fit to the WMAP TT power spectrum data and the $\chi^2$ is for this
fit with 893 degrees of freedom.}
\label{tab:model}
\end{table}

We first pick the physical matter density
$\omega_{\rm mat}\equiv\Omega_{\rm mat}h^2$, the (comoving) distance
to last scattering, $d_{\rm ls}$, and a (constant) equation of state of
the dark energy, $w$.  We approximate the redshift of last scattering as
$z=1080$, ignoring the slight matter and baryon density dependence.
This then allows us to solve for the Hubble constant, $h$, and thus
$\Omega_{\rm mat}$ and $\Omega_{\rm de}$ assuming spatial flatness.
Because they are reasonably well known, compared to some other parameters,
we fix $\omega_{\rm mat}=0.145$ and $d_{\rm ls}=13.7\,$Gpc for all of the
models in our grid.  These numbers are close to the best fit for a recent
analysis of WMAP and SDSS data \citep{Tegmark}.

The next step is to specify the other model parameters, for example the
optical depth to Thomson scattering, $\tau$, the spectral index, $n$
and the baryon density $\omega_b$.
(The optical depth enters primarily through its effect on the normalization
of the power spectrum when we fit to WMAP.)
We currently hold the number of light neutrinos fixed, and include no
massive neutrinos.  We deal with pure power-law spectra with no running
spectral index.
Our primary variations are in the optical depth, the spectral index and
the equation of state of the dark energy (see Table \ref{tab:model}).
The first two affect the amplitude and shape of the density or potential
fluctuations at late times while $w$ is a parameter of great interest to
the cosmology community.
We use a fixed equation of state in this initial survey, though it is easy
to include any known functional form in the future.

\begin{figure}
\begin{center}
\resizebox{3.5in}{!}{\includegraphics{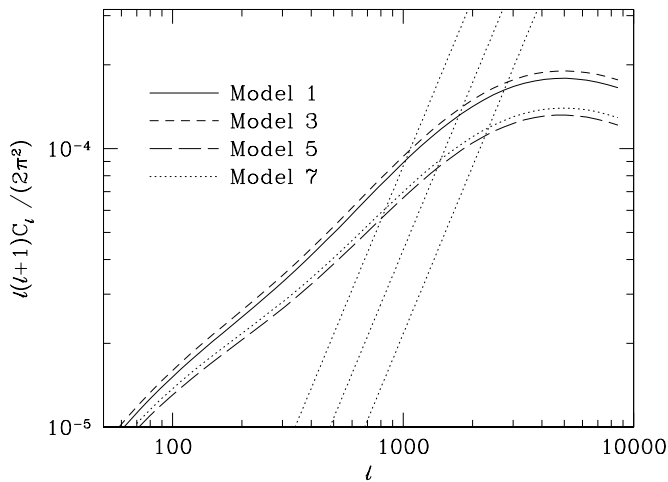}}
\end{center}
\caption{The angular power spectra predicted for our 4 cosmological
models for sources at $z_s\equiv 1$.  We assumed the Born and Limber
approximations, and used an ansatz for the non-linear mass power
spectrum to make these predictions.  The dotted lines rising steeply
to top right are the shot-noise power spectra for $\gamma_{\rm rms}=0.4$
and $\bar{n}_{\rm gal}=25$, 50 and $100\,$arcmin${}^{-2}$.}
\label{fig:lcl}
\end{figure}

For a specified model we compute the CMB anisotropy spectrum using v4.5
of {\sl CMBFAST\/} \citep{cmbfast}. 
The likelihood software provided by the WMAP team \citep{WMAP1,WMAP2,WMAP3} 
and the
known anisotropy spectrum are then used to find the best fitting normalization
of the power spectrum, $\sigma_8$, which is then converted into internal
code units for the N-body simulations.
In this way a file of model parameters, shown in Table \ref{tab:model}, is
built up.
The lensing convergence angular power spectra for our 4 models, assuming the
Born and Limber approximations, are shown in Fig.~\ref{fig:lcl}.
Note that the spectra for each pair of models have similar shapes, differing
primarily in amplitude.
In each case the model with the most distant sources has the highest
amplitude, even though it has a slightly lower matter power spectrum
amplitude, $\sigma_8$.
There will also be differences in the higher order moments and the source
redshift dependence of the power spectra.

\subsection{N-body simulations} \label{sec:nbody}

Another Perl program reads this cosmological parameter file and generates
a directory structure containing the requisite auxiliary files for the
N-body simulation, plus scripts for generating the initial conditions and
running the N-body code on the IBM-SP {\sl Seaborg\/} at NERSC.

All simulations require choices to be made about box size, resolution and
sampling.  To minimize numerical artifacts we want to run the largest box
with the highest dynamic range consistent with the computational resources.
Since this is a pilot program we have chosen simulations which use $256^3$
particles in a $200\,h^{-1}$Mpc box.
As we shall discuss in \S\ref{sec:numerics}, this is expected to be
sufficient for our purposes now, but will need to be improved upon in
the future.  As further resources become available we can increase the
volume simulated and/or decrease the force softening.

The initial conditions are generated by displacing equal mass particles from
a ``fuzzy'' Cartesian grid at $z=50$ using the Zel'dovich approximation.
The initial fluctuations are Gaussian, however we choose from a larger
number of initial conditions those whose low few $k$-modes are close to
the mean.  This avoids running any simulations which are significant outliers
and may skew statistics for a small number of runs.
For each cosmological model we choose two independent initial conditions.
Ideally we would run more realizations of each model to increase the volume
simulated, but with two runs we will be able to simulate several hundred
square degrees which is enough for now.

The {\sl TreePM\/} code, described in \cite{TreePM}, is used to
evolve the particles to $z=0$ with the phase space information output
at selected times after $z=3$.
The outputs are spaced equally in conformal time, with a spacing equal
to the time taken for light to travel $50\,h^{-1}$Mpc (comoving).
This spacing is small enough to obtain good lensing maps -- a halo with
``typical'' velocity $300\,$km/s will move only $50\,h^{-1}$kpc between
outputs, and its departure from straight line motion will be small.  
Each simulation gives between 80 and 90 phase space dumps, depending on
the cosmology, each of $\sim500\,$MB.
The gravitational force softening is of a spline form, with a
``Plummer equivalent'' softening length of about $30\,h^{-1}$kpc, again
small enough for our purposes.

Since the box size is only $200\,h^{-1}$Mpc, leading to a relatively coarse
sampling in $k$, we generate the initial conditions from a fit to the
transfer function which has the baryon oscillations ``smoothed out''.
For now we use the fit published by \cite{EisHu}.
This fit was made to an older version of {\sl CMBFAST}, and since that time
improvements in the code and the physics \citep{SSWZ} have changed the
predictions for $T(k)$ slightly.  The differences are small, and for the
purposes of testing and calibrating algorithms it is enough to know the
input $T(k)$ precisely, but in the future we will improve this aspect.

\subsection{Making lens maps} \label{sec:lensing}

We make two sets of maps.  The first, which we shall call the Born series,
assumes that the convergence field, $\kappa$, is simply the integral along
the line of sight of the density field weighted by a simple kernel
\begin{equation}
  \kappa \simeq {3\over 2}H_0^2\Omega_{\rm mat} \int d\chi
  \ g(\chi) {\delta\over a}
\end{equation}
where $\delta$ is the overdensity, $a$ is the scale-factor, $\chi$ is the
comoving distance and $g(\chi)$ is the lensing weight
\begin{equation}
  g(\chi) \equiv \int_\chi^\infty d\chi_s p(\chi_s)
    {\chi(\chi_s-\chi)\over\chi_s}
\end{equation}
for sources with distribution $p(\chi_s)$ normalized to $\int dp=1$.
This set of maps has the advantage of not requiring a Fourier transform in
its construction, providing higher resolution for a given pixelization.  Its
principal disadvantage is the approximate nature of the calculation.

We make several sets of 16 quasi-independent $3^\circ\times 3^\circ$ maps
with $1024^2$ pixels.
The first set of maps has sources fixed at $z_s\equiv 1$,
while the second set uses a source distribution of the form
\citep{Brainerd}
\begin{equation}
  {dp\over dz_s} \propto z_s^2 \exp\left[ -(z_s/z_0)^{3/2} \right]
\label{eqn:dpdz}
\end{equation}
for $z_0=2/3$ and $1$.
For this distribution $\langle z\rangle=\Gamma({8\over 3})z_0 \simeq 1.5\,z_0$,
but because there is more of the lensing weight at higher $z$ the lensing
signal is smaller for this distribution than a $\delta$-function distribution
with $z_s=1.5\,z_0$.

The second set of maps comes from a full multi-plane ray tracing algorithm
as described in \cite{ValWhi}.
We make maps of the shear components, $\gamma_i$, and the convergence,
$\kappa$, at a range of source redshifts from $z\sim 0$ to $3$ in steps
of $50\,h^{-1}$Mpc.
Again we make 16 maps per simulation, choosing different positions and
orientations for the `observer'.
In each case a $2048^2$ grid of rays subtending a field of view of $3^\circ$
is traced through the simulation, with the required Fourier transforms being
done on a $2048^2$ grid.  All assignments to and from the Fourier grid are
done with cloud-in-cell (CIC) assignment \citep{HocEas}.

\begin{figure}
\begin{center}
\resizebox{3.5in}{!}{\includegraphics{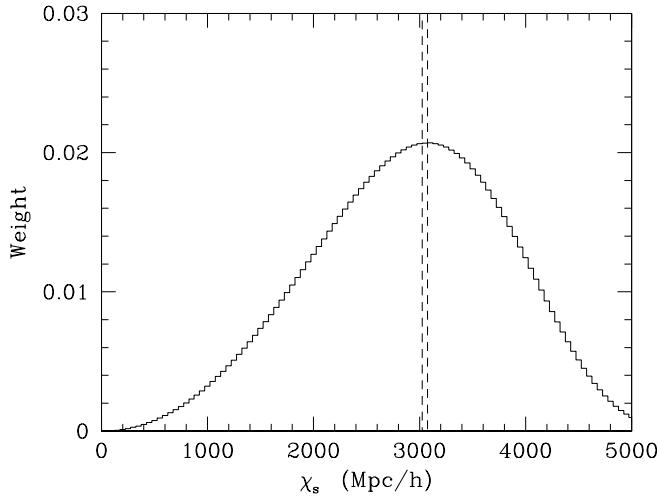}}
\end{center}
\caption{An example of the weights to be applied to the source planes of
model 1 in order to construct maps appropriate to Eq.~(\ref{eqn:dpdz})
with $z_0=1$ (solid) and a trivial $z_s\equiv 1.5$ (dashed).  The planes
are spaced by $\Delta\chi=50\,h^{-1}$Mpc.
The weight for $z_s\equiv 1.5$ is unity in a single bin.
The area under the solid histogram is also unity, but 4\% of this area
lies beyond $z=3$ where our simulations stop.}
\label{fig:lenswt}
\end{figure}

The two shear components and the convergence are output at each source plane.
{}From this a map of the (measurable) reduced shear, $\gamma_i/(1-\kappa)$,
for any source distribution can be computed.  For a distribution $dp/dz_s$
the weight given to source plane $j$ is
\begin{equation}
  w_j = \left. {dp\over dz_s}\right|_j H(z_j) \Delta\chi
\end{equation}
where $\Delta\chi=50\,h^{-1}$Mpc is the spacing between outputs.
The final shear at each point is then $\gamma_i=\sum_j w_j\gamma^{(j)}_i$
with $\gamma^{(j)}_i$ the $i$th component of the shear measured on plane $j$.
We show an example of the weighting factors for $z_0=1$ in
Fig.~\ref{fig:lenswt}, along with the (trivial) weight for a $\delta$-function
source distribution at $z_s\equiv 1.5$.
The integral over $\chi_s$ is extremely well approximated by a sum over the
output planes.
These maps can thus be combined to produce lensing maps for a wide range of
source distributions, allowing source tomography to be used.
It is straightforward to introduce mock `galaxies', with an appropriate
distribution of intrinsic ellipticities, at this stage and produce mock
catalogues.  On each $\chi_s$ slice, the angular positions of the galaxies
can be made to trace the dark matter with some fidelity, giving rise to
source clustering effects.  Reasonable choices produce angular correlations
and redshift distributions for the `galaxies' which match current data well.
However the parameter space of such catalogues is large, so we have chosen
instead to make available simple co-added shear maps which can be
post-processed with a wide variety of simulated galaxy properties to make
shear maps.
To save space we provide the maps downsampled from $2048^2$ to $1024^2$.

\subsection{Numerics} \label{sec:numerics}

\begin{figure}
\begin{center}
\resizebox{3.5in}{!}{\includegraphics{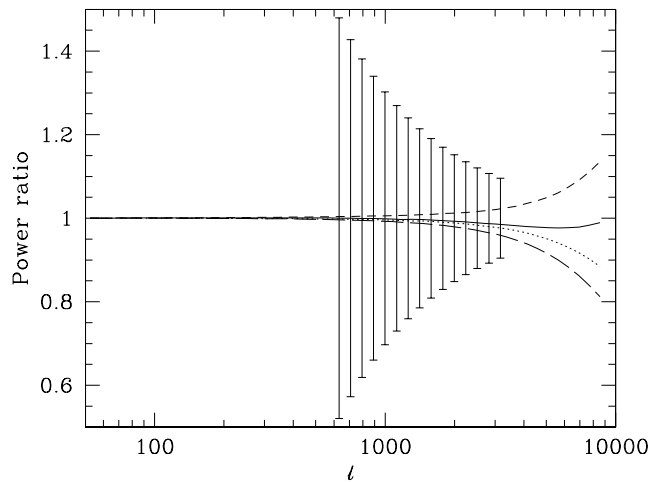}}
\end{center}
\caption{Resolution effects in the power spectra of the maps, compared
to the sample variance in a $3^\circ\times 3^\circ$ map.
The $y$-axis shows the ratio of lensing power spectra to a fiducial model
prediction.  All calculations are done with the 3D mass power spectrum fit
of \protect\cite{PeaSmi}, the Limber approximation and the model of
\protect\cite{ValWhi}.
The dotted line shows the effect of the finite force resolution in the
N-body simulation, the short-dashed line adds the finite mass resolution.
The long-dashed line shows the effect of the FT grid in the lensing
algorithm while the solid line combines all of these effects.
The error bars indicate the sample variance expected in a
$3^\circ\times 3^\circ$ field.}
\label{fig:numerics}
\end{figure}

To understand the fidelity of the map making procedure we need to
account for the finite volume, force and mass resolution of the N-body
simulation and the effect of the finite sampling during ray tracing.
The comoving distance to our furthest sources is $3\,$Gpc, at which
distance our box subtends $3.8^\circ$.  The volume of space probed by
a lensing field is $\sim 10^7\,{\rm Mpc}^3$, comparable to the volume
simulated in each run.

In \cite{ValWhi}, we developed an analytic model to estimate the
effects of finite force and mass resolution and the Fourier grid used in
the ray tracing.
This model was based on modifying the 3D dark matter power spectrum that
appears in the Limber integral for $C_\ell$ in a way designed to fit the
effects seen in a series of N-body simulations.  This allows us to compare
a given numerical configuration to a ``perfect'' model.
Fig.~\ref{fig:numerics} illustrates these effects individually for model 1
of Table \ref{tab:model} with sources at $z_s\equiv 1$.  For simplicity we
use the power spectrum fit of \cite{PeaSmi} in computing these ratios.
Note that the suppression of power by the finite force resolution and
FT grid is partly countered by the artificial increase in power due to
shot-noise.
Based on Fig.~\ref{fig:numerics} we expect our maps to be accurate to
several percent, in the 2-point function, for $\ell\le 3000$ and even
higher in some situations where the cancellation is accurate.

While the 16 maps per model are not fully independent, they do sample
different regions of the box in different stages of evolution and with
different projection effects.
The fraction of the volume traced by the rays in the field of view,
weighted by the contribution to the variance of the convergence from
each box, is between 10-20\% (for $100<\ell<3000$).
This forms a rough upper limit to the degree of correlation between the
maps.
Including the two independent boxes per model we have simulated several
hundred square degrees of sky per cosmology, which is close to the amount
of observational data currently existing.
Extending this to more sky is a straightforward exercise that simply requires
more computing resources.

\section{Results} \label{sec:results}

\begin{figure}
\begin{center}
\resizebox{3.5in}{!}{\includegraphics{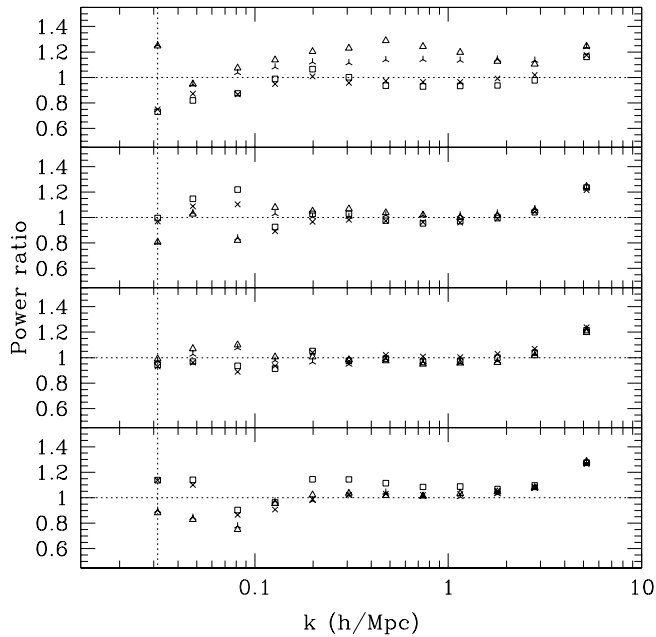}}
\end{center}
\caption{A comparison of the 3D mass power spectrum with the fitting function
of \protect\cite{PeaSmi} at selected redshifts.  The $y$-axis shows the
ratio of the numerical to the semi-analytic result, with the two independent
runs per cosmology and two different redshifts shown as different symbol
types in each panel.  Open symbols are for $z=0$ and starred symbols are
for $z\simeq 1$.  The top panel is models 1 and 2, then 3 and 4 and so on.
The vertical dotted line marks the fundamental mode of the simulations.}
\label{fig:pk}
\end{figure}

These simulations and lensing maps can be used for a variety of
purposes.  Here we simply show some preliminary analysis to orient the
reader.  The real uses of these maps will be in testing analysis algorithms
and new ideas.

Before turning to the lensing maps let us consider the underlying mass
distribution.
Fig.~\ref{fig:pk} shows the 3D matter power spectrum from our 8 boxes at
$z=1$ and $z=0$ compared to the semi-analytic model of \cite{PeaSmi}.
The agreement is good for both the $\Lambda$CDM models and the models
with equation of state $w=-0.8$, as expected.  On the scales simulated
the dark energy only enters through the Hubble parameter in the drag term,
slowing the growth of large-scale fluctuations when the dark energy comes
to dominate the expansion.  Since the fit of \cite{PeaSmi} has been
tested for open and $\Lambda$CDM models it is no surprise that it works
equally well for $w=-0.8$.
The disagreement at the highest $k$ might be due to numerical artifacts
in the N-body simulation or in the computation of the power spectrum,
however these points are significantly above both the shot-noise and force
smoothing limits.
We find similar behavior in a number of higher resolution simulations we
have analyzed, and a similar level of disagreement can be found in some of
the figures of \cite{PeaSmi} for CDM models and may reflect inherent
inaccuracy in the fitting function.
A qualitatively similar level of agreement was also found between
particle-mesh simulations and the semi-analytic model by \cite{IHMS},
though they were unable to probe very high $k$-modes due to limited force
resolution.

As a further check on the N-body code we investigated the scaling of the
power spectrum of an $n=-1$ self-similar model with the same numerical
parameters as the models described.  This suggests that the power spectrum
should be accurate to a few percent on the range of scales of interest.

\begin{figure}
\begin{center}
\resizebox{3.5in}{!}{\includegraphics{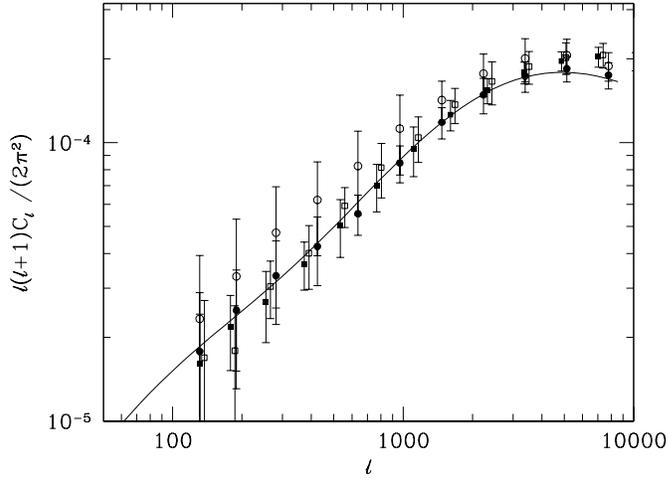}}
\end{center}
\caption{The angular power spectra predicted for model 1 with $z_s\equiv 1$.
The symbols with error bars are the mean and variance of the spectra from
16 maps, each $3^\circ\times 3^\circ$, from runs 1 and 2.  The points have
been offset slightly (horizontally) for clarity.  The open and filled squares
are the results from our ``Born'' series while the open and filled circles are
for the full ray tracing.}
\label{fig:lcl12}
\end{figure}

\begin{figure}
\begin{center}
\resizebox{3.5in}{!}{\includegraphics{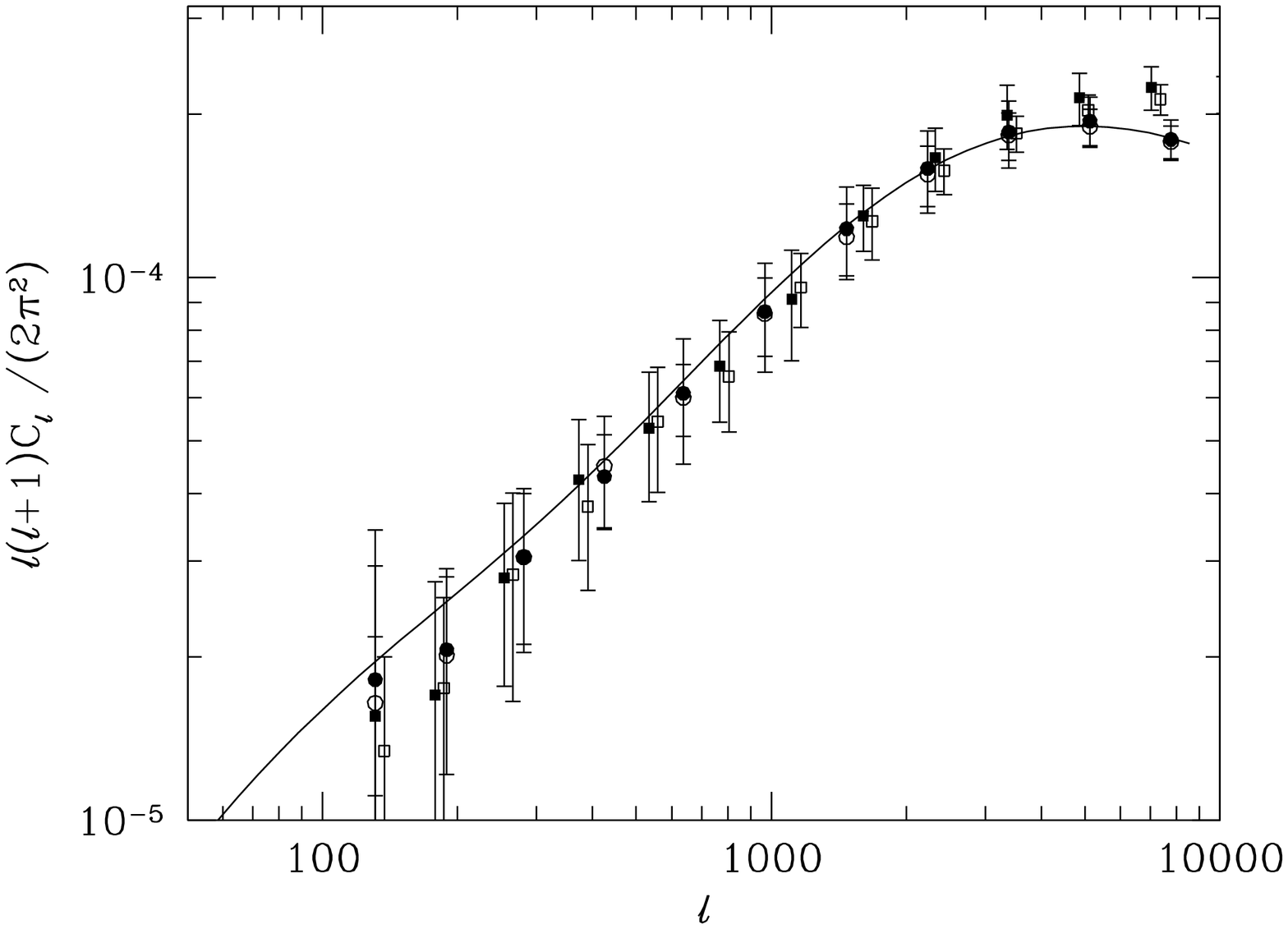}}
\end{center}
\caption{The angular power spectra predicted for model 3 with $z_s\equiv 1$.
The symbols with error bars are the mean and variance of the spectra from
16 maps, each $3^\circ\times 3^\circ$, from runs 3 and 4.  The points have
been offset slightly (horizontally) for clarity.  The open and filled squares
are the results from our ``Born'' series while the open and filled circles are
for the full ray tracing.}
\label{fig:lcl34}
\end{figure}

\begin{figure}
\begin{center}
\resizebox{3.5in}{!}{\includegraphics{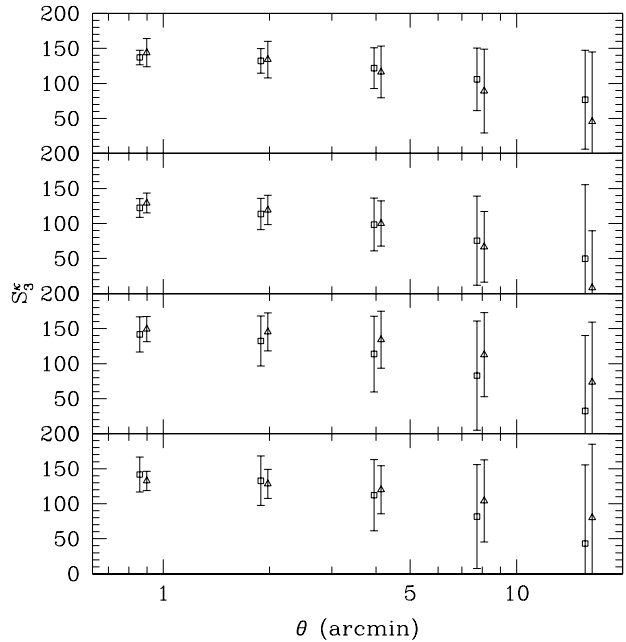}}
\end{center}
\caption{The skewness of the convergence, $\kappa$, for models 1-8 for
sources at $z_s\equiv 1$.
The symbols with error bars are the mean and variance of $S_3$ from 16
`Born' maps, each $3^\circ\times 3^\circ$.  The results from the ray tracing
are very similar and are omitted.
The points have been offset slightly (horizontally) for clarity.}
\label{fig:s3}
\end{figure}

The $\kappa$ angular power spectrum produced from 16 maps each for models
1 \& 2 is shown in Fig.~\ref{fig:lcl12}.  For this comparison we show maps
made assuming the Born approximation and with full ray-tracing, both for
sources fixed at $z_s\equiv 1$.  The agreement is generally good, with the
variance at low-$\ell$ expected from sampling 16 $3^\circ\times 3^\circ$
fields.
The open circles, corresponding to model 2, are slightly higher than the
semi-analytic model at intermediate $\ell$, as expected from the excess
power seen in the top panel of Fig.~\ref{fig:pk} (the triangles) at
intermediate $k$.
At the highest $\ell$ the `Born' maps show a slight excess power which could
be due to shot-noise in the simulations or to a breakdown in some of the
approximations we have made.  Alternatively the semi-analytic model and the
ray tracing could be underestimating the power on these scales, as suggested
by Figs.~\ref{fig:numerics} and \ref{fig:pk}.
In measurements these scales would be significantly affected by finite galaxy
ellipticities and densities, so this disagreement is likely unimportant.
Fig.~\ref{fig:lcl34} shows the same comparison for models 3 and 4.  The
power spectra for the remaining four models are very similar and we do not
show them explicitly here.

\begin{figure}
\begin{center}
\resizebox{3.5in}{!}{\includegraphics{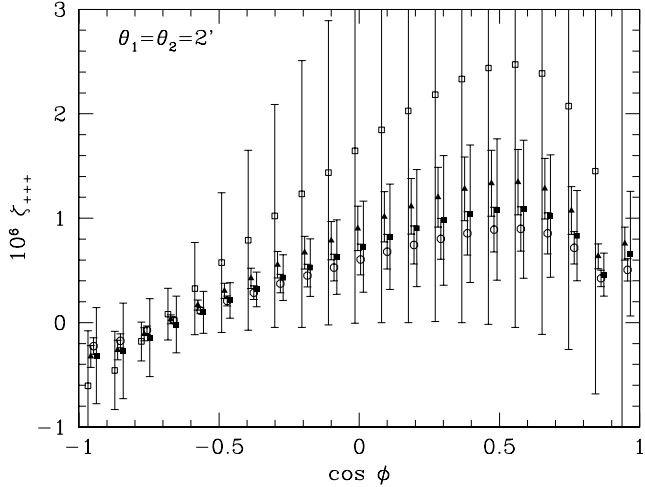}}
\end{center}
\caption{The shear 3-point function, $\zeta_{+++}$, for isoceles triangles
with $\theta_1=\theta_2=2'$ as a function of the cosine of the included
angle.  Different symbol types represent the mean and standard deviation of
32 maps, each $3^\circ\times 3^\circ$, for the 4 cosmologies simulated:
models 1 \& 2 (open squares), 3 \& 4 (filled triangles),
5 \& 6 (open circles), 7 \& 8 (filled squares).  Points have been offset
slightly horizontally for clarity.  In each case we assume $z\equiv 1$.}
\label{fig:tpt}
\end{figure}

\begin{figure}
\begin{center}
\resizebox{3.5in}{!}{\includegraphics{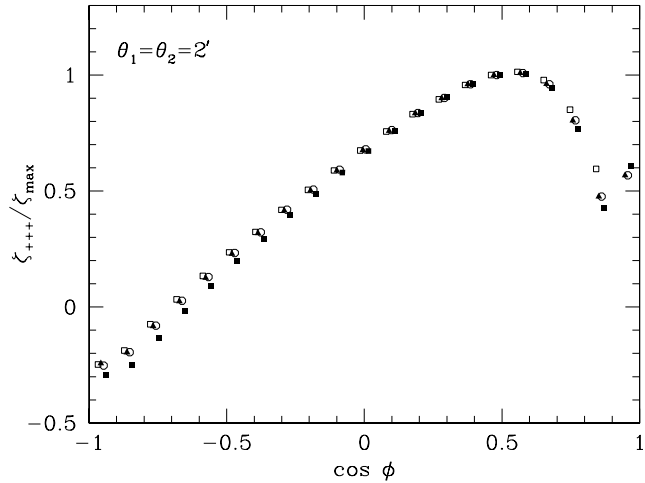}}
\end{center}
\caption{The shear 3-point function, $\zeta_{+++}$, for isoceles triangles
with $\theta_1=\theta_2=2'$ as for Fig.~\protect\ref{fig:tpt}, except
scaled to unity at maxmimum.  Note that the shape of the curves is very
similar among all of the models.  Symbol types are as in
Fig.~\protect\ref{fig:tpt}.}
\label{fig:tpt_scaled}
\end{figure}

The maps are non-Gaussian, as expected, and we provide the skewness
\begin{equation}
  S_3^\kappa \equiv {\langle\kappa^3\rangle\over\langle\kappa^2\rangle^2}
\end{equation}
for each of the models in Fig.~\ref{fig:s3}.  The $\kappa$ field is
smoothed by a boxcar of side length $\theta$ before the moments are
computed.  The plot shows the average and the variance across the 16
maps as before.
As noted previously \citep{WhiHu,ValWhi} the skewness suffers from a
large sample variance.

We also show the shear 3-point function for isoceles triangles of side
length $\theta_1=\theta_2=2'$ in Fig.~\ref{fig:tpt}.
Here we plot $10^6\zeta_{+++}$ as a function of the cosine of the
included angle for the 32 fields, each $3^\circ\times 3^\circ$, simulated
for each cosmology assuming the sources are at $z_s\equiv 1$.
The points are the mean of the 32 fields, and the error bars represent the
variance.
The highest 3-point function is for models 1 \& 2, which have the highest
normalization ($\sigma_8$) of all of the models.  Models 3 \& 4 have the
next largest amplitude, and the second highest normalization.
For models 5 \& 6 the effect of a higher normalization isn't enough to
overcome the less negative $w$ of models 7 \& 8, which have a slightly
higher amplitude.
Notice the large scatter from field to field, but the familiar shape
\citep{TakJai02}.
Apart from an amplitude change the shapes of the curves are extremely
similar, as shown in Fig.~\ref{fig:tpt_scaled}.
We find a similar behavior for other configurations we have checked, which
has interesting implications for the signal encoded in the non-Gaussianity
of the maps and how best to extract it.
We also note that the peak of $\zeta_{+++}$ appears to be shifted slightly
{}from the equilateral configuration, due to the asphericity and substructure
of the dark matter halos \citep{HoWhi,DolJaiTak}.

\section{Conclusions} \label{sec:conclusions}

Gravitational lensing has made rapid advances in recent years, and
observations are now pushing the limits of existing theoretical models.
If lensing is to fulfill its promise as a precision tool, theorists
need to improve the predictions of cosmological models for the $n$-point
correlation functions.  An important step in this program is the creation
of grids of N-body models of sufficient resolution and sampling to enable
accurate simulation of weak lensing.
This paper describes a first step in this direction.

Based on analytic arguments developed in \cite{ValWhi} we believe
that simulations with $256^3$ particles in boxes of side $200\,h^{-1}$Mpc
are sufficient to produce lensing maps accurate to $\ell\simeq 2000-3000$
or scales of a few arcminutes.
To achieve good convergence in the multi-plane ray-tracing algorithm that
we use requires time dumps spaced equally in conformal time with a spacing
close to $50\,h^{-1}$Mpc.

The total computational cost of this project was very modest.
The major human cost is the time spent managing the disk space,
archiving and retrieving the data to ensure that the total usage
stays below quota.
The simulations each took 2000-3000 time steps, requiring slightly under
one day each on 32 processors of the IBM-SP {\sl Seaborg\/} at NERSC,
i.e.~between 640 and 720 CPU hours.
The phase space data required $\sim 40\,$GB per model while the ray tracing
simulations required around $100\,$GB of intermediate storage and 100 CPU
hours per model.  The disk usage is driven by the need for all the phase
space data and the $80-90$ high resolution source planes for each of 16 maps.
The final stacked and downsampled maps are $<1\,$GB per model and source
distribution.
The entire process was monitored and controlled by scripts, making it easy
to extend the grid as more resources become available.

We have made the raw maps, along with some auxiliary data products, freely
available to the community at http://mwhite.berkeley.edu/ in the hope that
they will be useful in taking the next step.

M.W. thanks Tzu-Ching Chang, Dragan Huterer, Bhuvnesh Jain, Jason Rhodes
and Masahiro Takada for helpful comments on an earlier draft.
The simulations used here were performed on the IBM-SP at the National
Energy Research Scientific Computing Center.
This research was supported by the NSF and NASA.

\end{document}